\documentclass[pra, amsmath,amssymb,aps,10pt,twocolumn,superscriptaddress]{revtex4-1}
\usepackage{graphicx}
\usepackage[usenames, dvipsnames]{color}
\usepackage{dcolumn}
\usepackage{bm}
\usepackage{color}
\usepackage[colorlinks,bookmarks=false,citecolor=blue,linkcolor=blue,urlcolor=black]{hyperref}

\begin{document}

\preprint{APS/123-QED}

\title{Strong-field ionization of atoms with $p^3$ valence shell: Two versus three active electrons}
\author{Dmitry K. Efimov} \email{dmitry.efimov@uj.edu.pl}

\affiliation{Institute of Theoretical Physics, Jagiellonian University in Krakow, \L{}ojasiewicza 11, 30-348 Krak\'ow, Poland}
\author{Jakub S. Prauzner-Bechcicki}
\affiliation{Instytut Fizyki imienia Mariana Smoluchowskiego, Uniwersytet Jagiello\'nski, \L{}ojasiewicza 11, 30-348 Krak\'ow, Poland}
\author{Jakub Zakrzewski}
\affiliation{Institute of Theoretical Physics, Jagiellonian University in Krakow, \L{}ojasiewicza 11, 30-348 Krak\'ow, Poland}
\affiliation{Mark Kac Complex Systems Research Center, Jagiellonian University in Krakow, \L{}ojasiewicza 11, 30-348 Krak\'ow, Poland}
\date{\today}

\begin{abstract}
	For a model atom with the $p^3$ valence shell we construct consistent three- and two-active electrons models enabling their direct comparison. Within these models, we study the influence of the third active electron on the double ionization yield in strong femtosecond laser fields. We reveal proportionality between double ionization signals obtained with both models in the field intensity region where non-sequential ionization dominates. 
	We derive analytically a correspondence rule connecting the double ionization yields obtained within the three- and two-active electrons models.
	
\end{abstract}

\maketitle

	\section{\label{sec:introduction}Introduction}

	Physics of many-body systems interacting with strong, time-dependent fields is a broad subject of current research \cite{Bauer17,Grossmann18}. This area of research includes studies on multiple-electron atoms \cite{Wahyutama19,Chen19,Prager01,Kondo93,Bergues12,Lotstedt16}, molecules \cite{Nisoli17,Majety17,Winney18,Palacios19,Telnov09,chattopadhyay2018electron} and condensed matter systems \cite{Ghimire19,Navarrete19,Yu19,Imai19,Lu19}. The reaction of these systems to a strong external field  depends both on the interaction of electrons with the field and on the interaction of electrons with each other. Collective and correlated processes not only take place, but even dominate the response of the system in some specific regimes of field amplitudes. Such a response, in turn, can either be characteristic to many-electron systems and their complex structure or manifest itself in observables that exist even in simpler systems (one- or two-electron). Charge migration \cite{Nisoli17,Mauger18,Yuan18,Worner17} is an example of the former kind of response and high-harmonic generation (HHG)~\cite{Lewenstein94,Antoine96,Silva18,Masin18,Sukiasyan09,Smirnova09,Ivanov14multielectron,Li19,Tikhomirov17,Amini_2019}, single and double ionization~\cite{Faria11,Walker94,Colgan13,Fittinghoff92,Weber00,watsonPRA97,Goreslavskii01,Moshammer02,Chen17,becker2012theories,Pfeiffer11a} are examples of the latter.
	
	The simplest system, in a context of ionization, in which correlated processes take place is a two-electron atom. The strong-field physics of that kind of systems is fairly well understood~\cite{Faria11,becker2012theories,Becker00,Bergues15,haan2002,Liu99,panfili01,Haan07,Pfeiffer11a,ho2005classical}. Furthermore, the response of the two-electron atom to an external strong laser field is solvable numerically in full dimensional space~\cite{dundas99,parker1998intense,feist2008nonsequential,djiokap2017kinematical,Hu13}. However, the addition of only one more electron complicates the system so that it causes serious difficulties in studying it. First approach in treating strong-field ionization of three-electron atoms used the classical description~\cite{ho2006plane,Ho07,sacha2001triple,Emmanouilidou08b,Emmanouilidou06}. Yet, such approaches cannot account for electron spins, that impose significant restrictions on the symmetry of the wave function in realistic systems~\cite{Ruiz05}. Still, classical analysis~\cite{sacha2001triple} helps in developing simplified quantum models in which spin-dependent effects can be studied~\cite{Rapp14,Thiede18,Efimov19,Colgan13}.
	
	Due to the rising complexity of treating systems with more than two electrons, one naturally seeks for the possibility of reduction of a complex system to simpler system or set of subsystems. The simplified systems are expected to preserve the key elements of the dynamical response.
	If such a reduction is successful, observables of the complex system could be expressed as a combination of observables of the simpler system or subsystems.
	A good example can be found in \cite{Bray19}, where a HHG spectrum of a many-electron Xenon is represented by a spectrum generated with single electron time-dependent Schr\"odinger equation (TDSE) and then multiplied by a ratio of photoionization cross sections for different ionization channels of Xe.
	
	Not always such a solution can be found. In our previous work \cite{Efimov19} we have studied a class of atoms with $ns^2np^1$ electrons forming the outer shell, in which case a reduction to two-electron subsystems was in vain. The double ionization yields (DIYs) could not be reproduced by a combination of DIYs obtained with use of two-electron subsystems. The configuration of the outer shell electrons imposed symmetry constraints on a three-electron wavefunction. Two of $ns^2np^1$ electrons have the same spin thus spacial three-electron wavefunction is antisymmetric with respect to exchange of one pair of electrons and symmetric with respect to exchange of two other pairs. Consequently, in two-electron subsystems{,} one has to consider symmetric and antisymmetric wavefunctions, respectively. The final result of the work~\cite{Efimov19} can be rephrased as: in the case of $ns^2np^1$ electrons the full three-electron model cannot be represented by a combination of a two-electron models possessing different spatial symmetries. One can wonder, however, whether a correspondence in the sense od DIYs between the three- and two-electron models can be ever established.
	
	In order to investigate the above-stated question, we consider a system with three equivalent electrons, that is, electrons possessing the same spin. Consequently, the related two-active-electron model inevitably consists of two electrons with the same spin. In the language of spatial symmetries of wavefunction this means that the \textit{totally antisymmetric} three-electron wavefunction, i.e. the one that is antisymmetric with respect to any arbitrary electron pair interchange, can only be juxtaposed to the \textit{antisymmetric} two-electron wavefunction. 
	The spin effects are thus withdrawn from our consideration and the difference between the models' performances naturally reflects the difference between the two- and three- active-electrons models {\it per se}. 
	
	In the following, we shall examine the totally antisymmetric three- and two-electron wavefunctions in order to reveal a clear correspondence rule for models with different number of electrons taken into account.  As spins do not affect ionization dynamics, we drop all the spin indications and sum up ionization impacts from all electrons.
	The atoms with three valence electrons having the same spin can be divided into two groups. The first one consists of transition metal elements with a $d^3$ or $f^3$ valence shell, the another one is the chemically active 15th group atoms with a $p^3$ valence shell. As a model we have chosen atomic Nitrogen because of reasonable values of single and double ionization potentials, despite it can hardly been used as a target gas in the strong-field experiments. The values of ionization potentials define the range of laser intensities for which correlated processes are expected to be important. The magnitude of intensity, in turn, heavily affects the performance and applicability of the numerical algorithms used to study the proposed models.

	For such conditions, we have found that DIY as a function of field amplitude has a very similar shape in both three- and two-active electrons models, although considerably differs in magnitude. Such a result suggests the existence of a correspondence rule that allows for an unambiguous connection between the two models. To identify that rule we shall apply the quantitative rescattering (QRS) theory \cite{Chen19,Chen09}, in which a double ionization is reproduced with the application of three main factors: the returning electron wave packet, the differential rescattering cross section and the ionization rates from excited ionic states. As the first and the last are essentially the same for the two analyzed models, we propose that the correspondence rule is associated with properties of electronic rescattering cross sections.  
	
	The paper is organized as follows. We start with providing a  brief description of the three- and two- electron models together with a values of ionization potentials in Sec.~\ref{sec:models}. We further present in Section~\ref{sec:results} results of numerical simulations and then proceed with deriving a quantitative explanation of the observed DIY ratios. We close with a summary and conclusions in Section \ref{sec:conclusions}. Atomic units are used throughout this paper unless stated otherwise. For the sake of clarity, we note that 1 a.u. of energy is equal to 27.2 eV; at the same time 0.1 a.u. of electric field corresponds to $3.5\cdot 10^{14}$ W/cm$^2$ of laser intensity.

	\section{\label{sec:models}Models and methods}
	
	Several computational approaches to the problem formulated above could be adopted. With the current computational physics developments the application of time dependent density functional theory (TDDFT) \cite{Ulrich12} or time dependent multiconfiguration Hartree-Fock theory (TDMCHFT) \cite{Haxton11,brics2014nonsequential,Li17} could be a method of choice. Those methods, optimal in intermediate laser intensity regimes, in particular for  HHG spectrum simulations \cite{Li19} have problems when treating the dynamics of ionization and ionization yields for very strong field \cite{brics2014nonsequential,Chiril17}.  For these reasons we restrict to grid-based approach for reduced dimensionality model of the atom, an approach often used in the past particularly for linearly polarized laser field. Such reduced dimensionality models often serve as testbeds for checking the accuracy of more sophisticated approaches \cite{Lotstedt16} or are still used on their own (see e.g. \cite{Tikhomirov17,Eicke20}) also within TDDFT scheme \cite{Chiril17}.

In the traditional, most often used approach, each electron is allowed to move along one dimensional track along the polarization axis \cite{grobe93} and the Coulomb potential is modified with a soft-core parameter~\cite{su91}. Such an approach is known, however,  to overestimate the electron-electron repulsion and underestimate their correlation (as seen in its failure to reproduce correctly the characteristic knee feature associated with non sequential double ionization process \cite{becker2012theories}). Therefore we use the modified strategy that associates the one-dimensional electron tracks with the motion of saddles in the potential for quasi-static electric field of a variable amplitude. Such an ``adiabatic'' picture was proposed for double ionization almost twenty years ago for two \cite{sacha2001pathways} and three electrons \cite{sacha2001triple} and used successfully for two-electron \cite {prauzner2007time,prauzner2008quantum,eckhardt2010phase} and recently extended for three electrons \cite{Thiede18,Efimov19} problems. 
	
	\paragraph{\label{sec:3emod}Three-active electrons model.}  In our three-electron model tracks are inclined with respect to the laser polarization axis at the angle $\gamma$ ($\tan \gamma = \sqrt{2/3}$) and at the angle $\pi/6$ with respect to each other. Due to such a configuration we avoid overestimation of the electron-electron repulsion. The configuration is not arbitrary, it is identified on the basis of an adiabatic analysis of the ionization process~\cite{sacha2001triple}. In that analysis one finds efficient ionization channels by considering transition states, which are the saddles of the potential energy formed in the presence of the instantaneous static electric field. The saddles form a fixed configuration that moves along lines inclined at constant angle $\gamma$ with respect to the polarization axis and at constant angle $\pi/6$ with respect to each other as the field amplitude changes during the pulse~\cite{sacha2001triple}. The motion of electrons is then confined to those lines. 
	
	The Hamiltonian of three-electron system is 
	\begin{equation}
	H=\sum_{i=1}^3\frac{p_i^2}{2}+V(r_1,r_2,r_3)
	\label{ham3e}
	\end{equation}
	with 
	\begin{eqnarray}
	V(r_1,r_2,r_3)&=&-\sum_{i=1}^3\left(\frac{3}{\sqrt{r_i^2+\epsilon^2}} +\sqrt{\frac{2}{3}}F(t)r_i \right) \nonumber \\
	&+&\sum_{i,j=1 i<j}^3\frac{q_{ee}^2}{\sqrt{(r_i-r_j)^2+r_ir_j+\epsilon^2}},
	\label{pot3e_std}
	\end{eqnarray}
	where $r_i$ and $p_i$ correspond to the $i$-th electron's coordinate and momentum, respectively, and the field $F(t)$ is defined by $F(t) = -\partial A/\partial t$. 
	
	Because both the single and double ionization potentials of Nitrogen are defined uniquely, one can adjust the models to get the proper values of the potentials. For this purpose,
	we introduce in (\ref{pot3e_std}) an effective electron charge $q_{ee}$ to the electron-electron interaction term. This way, in both three- and two-electron models we have just two model parameters: the soft-core parameter $\epsilon$ and the effective electron-electron charge $q_{ee}$, that allow us to obtain the same single and double ionization potentials.  
	
	\paragraph{Two-active electrons model.}
	The two-active electrons model is built consistently from the three electrons model introduced above. Therefore, the electronic motion is restricted to one-dimensional tracks that form a plane and cross at angle $\pi/6$ as in the three electrons model. The electric field vector is forced to lie in that plane -- in contrast to the three electrons case, thus forming a different angle $\delta$ ($\cos (\delta) = \sqrt{3}/2$) with electronic axes. For the sake of comparison between discussed models, we impose the electric field operator geometrical prefactors to be the same and equal to $\sqrt{2/3}$, as introduced earlier in (\ref{pot3e_std}). The two-electron Hamiltonian then reads
	
	\begin{multline}
	H = \sum_{i=1}^{2} \left( \frac{p_i^2}{2} - \frac{Z}{\sqrt{r_i^2 +\epsilon^2}} +  \sqrt{\frac{2}{3}}F(t)r_i \right) + \\ \frac{q_{ee}^2}{\sqrt{(r_1-r_2)^2 + r_1 r_2 + \epsilon^2}},
	\label{ham2e}
	\end{multline}   
	where $Z$ is a nuclear charge, set either as $Z=2$ (neutral atom) or $Z=3$ (single ion). 
	
	\paragraph{The laser pulse.}	
	The laser pulse is defined by the vector potential
	\begin{equation}
	A(t) = \frac{F_0}{\omega_0} \sin^2 \left( \frac{\pi t}{T_p} \right) \sin(\omega_0 t), \quad 0<t<T_p.
	\label{pulse}
	\end{equation}
	The pulse parameters are: $F_0$ the field amplitude, $\omega_0$ the frequency and the pulse length $T_p = 2\pi n_c /\omega_0$ that is taken to be a multiple of the number of cycles $n_c$. In the following we set $\omega_0=0.06$ a.u. that corresponds to about 760 nm of laser wavelength and the pulse to $n_c=5$ cycles. The field amplitude is varied.
	
	\paragraph{Ionization potentials.} The values of soft-core parameters and effective electron-electron charges are taken to reproduce the experimental values of single and double ionization potentials for Nitrogen atom in both three- and two-electron models. For the first, $\epsilon=\sqrt{1.02}$ and $q_{ee} = \sqrt{0.5}$, while for the second $\epsilon=\sqrt{2.0}$ and $q_{ee} = \sqrt{0.3}$. The single ionization potential is then 0.52 a.u., the double ionization potential is 1.61 a.u. and the triple ionization potential (for three-electron model) is 3.92 a.u.
	
	\paragraph{Evolution.}
	For each of the models, TDSE is solved on a~spatial grid with the use of the split operator technique and Fast Fourier transform. The algorithms are described in details elsewhere~\cite{Thiede18,prauzner2008quantum,Efimov18}. Regardless of the model, the grid has 2048 points in each direction covering 400 a.u. of the physical coordinate space. Absorbing boundary conditions at edges of the integration box are used in a similar manner to~\cite{prauzner2008quantum}. The initial state is found by an imaginary time propagation in an appropriate symmetry subspace for a much smaller grid involving 512 points in each direction corresponding to 100 a.u.
	
	\section{\label{sec:results}Results and discussion}

	In the following we shall focus on the double ionization.
	To calculate double ionization yields we use a spacial criterion that we recall here in a nutshell, extended description can be found elsewhere~\citep{prauzner2008quantum,prauzner2007time,Thiede18,dundas99}.
	
	First of all, let us discuss the observables we obtain during evaluation of our numerical code. The coordination space is divided into the regions corresponding to the neutral (A), single ionized (S) and double ionized (D) atomic states (see Appendix \ref{App} for details of our approach).
	In a case of three electron model, there is also a region corresponding to a triply ionized state.
	The ionization yields are defined as integrated probability fluxes through borders of different regions.  	Such an approach allows one to numerically distinguish between two channels of double ionization: the direct double ionization and the time delayed double ionization.
	The direct double ionization is calculated as an integrated flux through the borders between the (A) and (D) regions and is assumed to describe processes in which both electrons leave the parent ion simultaneously. The dominant ingredient of that channel is the so called recollision induced ionization (RII) (that channel includes also the simultaneous tunneling of both electrons, a process, which is expected to be negligible). 
	The time delayed ionization (TDI) is calculated as an integrated flux through the borders of (S) and (D) regions.
	It accounts for  processes where electrons leave the parent ion in different instants of time.
	The spatial criterion for defining TDI inevitably puts into this channel both the sequential double ionization (SDI) and the recollision excitation with subsequent ionization (RESI). Thus pure SDI process cannot be resolved with the above-described method.
	
\begin{figure}
		\includegraphics[width=1.0\linewidth]{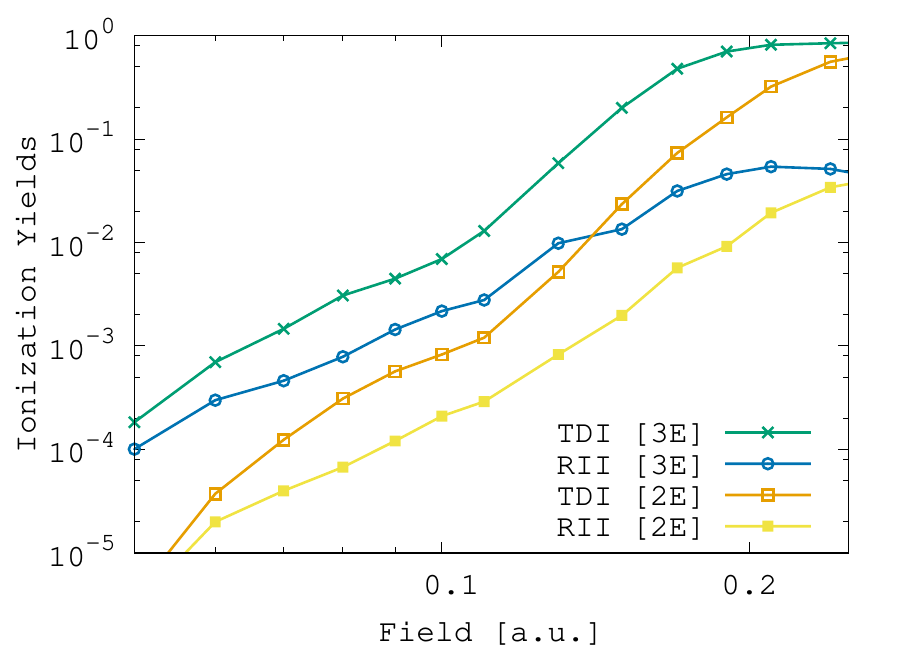}
		\caption{(Color online) Double ionization yields as a function of electric field amplitude resolved for different ionization channels of the three-active electrons model [3E] (\ref{ham3e}) and of the two-active electrons model [2E] (\ref{ham2e}). The channels are denoted as time delayed ionization (TDI) and recollision-impact ionization (RII). 5-cycles long $\sin^2$-shaped pulse of frequency 0.06 a.u. (Eq.~(\ref{pulse})) has been used for simulations.}
		\label{fig1}
		\includegraphics[width=1.0\linewidth]{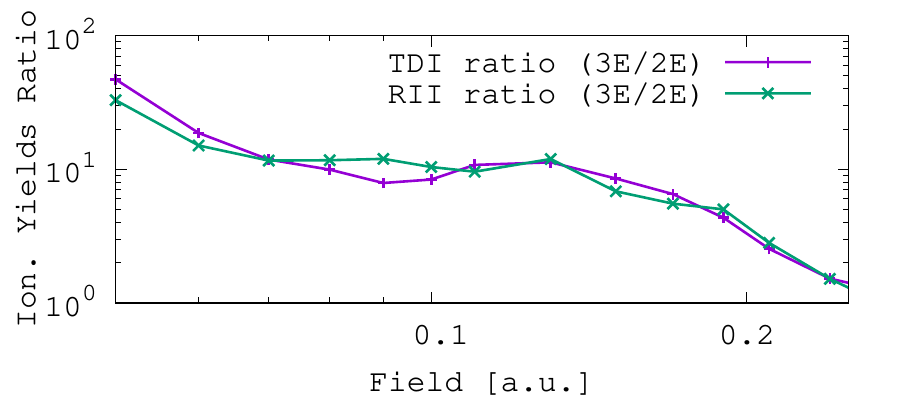}
		\caption{(Color online) Ratios of ionization yields obtained with the three-active-electrons and the two-active electrons models in TDI and RII  channels, respectively, as a function of electric field amplitude. The plot corresponds to the data in Fig.~\ref{fig1}.
		}
		\label{fig2}
	\end{figure}

	The dependencies of double ionization yields on the field amplitude for both three- and two-electron models are shown in Fig. \ref{fig1}.
	In each case, the RII and TDI channels are plotted separately.
	One can notice that the characteristic knee shape, the indicator of non-sequential processes, is barely visible.
	This is not to be unexpected, since in our models the electron-electron interaction term is modified with the effective electron-electron charge $q_{ee}$.
	Low effective electron-electron charges, $q_{ee}<1$, reduce the efficiency of electronic rescatterings and thus of all non-sequential processes in general.
	{It is worth mentioning that such a reduction of rescattering efficiency leading to a partial or full disappearance of the knee is not uncommon is strong-laser-field physics. In particular, it has been observed for atoms in circular polarization \cite{Lai20,Mancuso16}. Its dependences on field intensity \cite{Fu12}, frequency \cite{Chen17b} and type of species \cite{Kamor13} have been studied.}

	The much more interesting observation comes from the comparison of ionization yields in each of the channels obtained by three-electron and two-electron models. In Fig.~\ref{fig2} we present respective ratios for TDI and RII channels. Both ratios are nearly flat in the range of field amplitudes from 0.06 up to 0.2 (especially in the knee regime, i.e. $F\in [0.08,0.15]$) -- in each case 3E yield is one order of magnitude larger than its 2E counterpart. Finally, the TDI and RII ratios show almost  identical behavior, in the sense of shape and magnitude, in the whole range of analyzed field amplitudes.   
	
	The observed constant ratio of ionization yields in both channels suggests that the ratio of recollision cross sections for three- and two-electron models is field independent: while the full cross sections are field-dependent, their field dependence is uniform for all models. To prove our point we further reduce the ratio of ionization yields to the ratio of recollision cross sections and thereafter show that the latter is indeed field-independent. 	
	
	\subsection{The ratio of cross sections.}  Following the standard QRS theory \cite{Micheau09,Chen10} one can express the double ionization yield $P_{double}$ as a sum of the two ingredients, i.e. TDI and RII:
	\begin{align}
	&P_{double} = P_{TDI} + P_{RII} \label{gemini}\\
	&P_{TDI} = \int Y_{0}^{tunn}(t)\, dt + \sum_{n} Y_{n}^{exci}\int Y_{n}^{tunn}(t)\, dt,  \label{lory} \\
	&P_{RII} = \int dE \, Y_{E}^{RII},  \quad E>0
	\label{nightingale}
	\end{align}	
	where $Y_{n}^{tunn}$, $Y_{n}^{exci}$ and $Y_{E}^{RII}$ denote rates of tunneling ionization from the $n$-th excited state of an ion, collisional excitation of an ion to the $n$-th state, and RII with energy $E$ transferred from the rescattered electron to free the second electron, correspondingly. We are going to consider the yields of TDI and RII from Eqs. (\ref{lory}) and (\ref{nightingale}) separately.
	
	\paragraph{TDI yields.} For the field amplitudes from the knee regime, direct emission of the second electron from the ionic ground state ($n=0$) is negligible. The interaction with the recolliding electron is a must. Therefore, we can drop the first term of (\ref{lory}). However, for the saturation regime, i.e.  $F>0.2$, the omitted term becomes dominating and the ratio $P_{TDI}^{3\text{E}}/P_{TDI}^{2\text{E}}$ tends to unity -- the trend observed in the Fig. \ref{fig2} for high field amplitudes region. From now on we shall denote the respective model with upper indexes, $^{\text{(3E)}}$ for three- and $^{\text{(2E)}}$ for two-electron model. 
	
	For low and medium field amplitudes Eq.~\eqref{lory} reduces to: 
	\begin{multline}
	P_{TDI} \simeq \sum_{n} Y_{n}^{exci}\int Y_{n}^{tunn}(t)\, dt = \\ \sum_{n} \int \frac{d\sigma_n}{dp} W(p) \, dp  \int Y_{n}^{tunn}(t)\, dt,
	\label{whale}
	\end{multline}
	where $d\sigma_n/dp$ denotes the differential cross section of an excitation of an ion from the ground state to $n$-th state by an incident electron of momentum $p$. $W(p)$ is the recolliding electronic wave packet.
	
	The electronic wave packets are the same for two- and three-electron models as motion of the recolliding electron in each case is constrained to one dimension. Furthermore, the laser-induced ionization rates $Y_{n}^{tunn}(t)$ are the same because the ionization potentials are the same \cite{Majorosi18}. Therefore, the only model-dependent element of $P_{TDI}$ is $d\sigma_n/dp$.
	
\begin{figure}
		\includegraphics[width=1.0\linewidth]{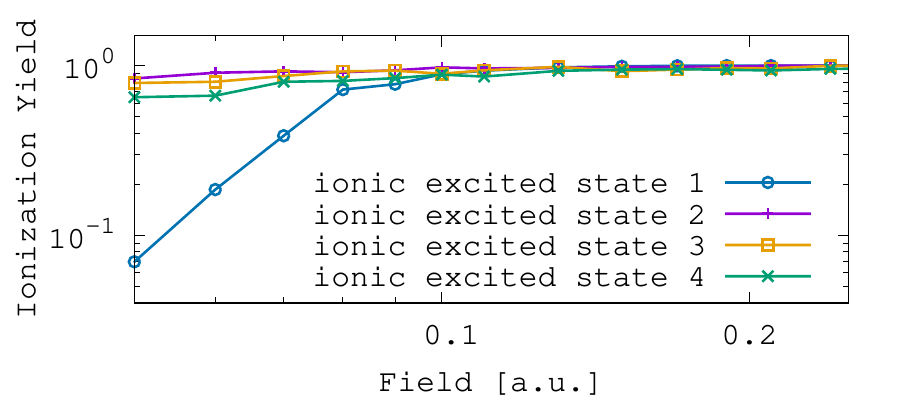}
		\caption{(Color online) Ionization yields of initially excited single ion. Curves correspond to different initial states. One-dimensional model with $\epsilon=\sqrt{2.0}$ is used. The eigenenergies of the first four excited states are --0.62, --0.40, --0.27 and --0.19 a.u., correspondingly. The field parameters are the same as in Fig.~\ref{fig1}.
		}
		\label{fig4}
	\end{figure}

	A direct evaluation of (\ref{whale}) is a complicated task.
	In the given regime of field intensities, however, a good approximation is to put $\int Y_{n}^{tunn}(t)\, dt \simeq 1$ for all excited states. Their relatively high eigenenergies provide the saturation of a laser-induced ionization for the field parameters used in the current research. That observation is nicely illustrated in Fig. \ref{fig4}, which shows ionization yields for first few excited states of an ion. 
	Now, TDI yield is expressed by:
	\begin{equation}
	P_{TDI} \simeq \sum_{n} \int \frac{d\sigma_n}{dp} W(p) \, dp = \int \left( \cfrac{d\sigma_{ex}}{dp}\right)W(p) \, dp,
	\label{full_excitation_cross_sec}
	\end{equation}
	where in the last step we introduced the full differential excitation cross section $d\sigma_{ex}/dp \equiv \sum_n d\sigma_n/dp$. In contrast to the set of $d\sigma_n/dp$, the full differential excitation cross section $d\sigma_{ex}/dp$ can serve as a good universal parameter characterizing efficiency of TDI.
	
	Let us assume that the differential excitation cross section can be factorized as
	\begin{equation}
	\cfrac{d\sigma_{ex}}{dp} = \tilde{\sigma}_{ex}f(p).
	\label{factorization_p}
	\end{equation}
	\noindent The first factor is dependent on model parameters only, $\epsilon$ and $q_{ee}$, while the second factor is only momentum-dependent. 
	Thanks to the factorization the TDI yields can be rewritten in a form:
	\begin{equation}
	P_{TDI} \simeq \tilde{\sigma}_{ex} \int f(p) W(p) \, dp.
	\end{equation}
	The integral is model independent thus we can express the analyzed TDI yields ration with ratio of $\tilde{\sigma}_{ex}$ that correspond to each model:
	\begin{equation}
	\frac{{P}_{TDI}^{3\text{E}}}{{P}_{TDI}^{2\text{E}}} \simeq  \frac{\tilde{\sigma}_{ex}^{3\text{E}}}{\tilde{\sigma}_{ex}^{2\text{E}}}.
	\label{TDI_ratio}
	\end{equation}

	\paragraph{RII yields.} Similarly to the case of TDI yields, the ratio $P_{RII}^{3\text{E}}/P_{RII}^{2\text{E}}$ saturates for $F>0.2$. The saturation directly follows from two facts: (i) that $P_{double}$ for either model is composed of two ingredients, $P_{TDI}$ and $P_{RII}$, see eq.~(\ref{gemini}) and (ii) that $P_{double}$ saturates for field amplitudes $F>0.2$.  
	
	To analyze the ratio of the RII yields for the fields $F<0.2$, let us rewrite eq.~(\ref{nightingale}) in a form:
	\begin{equation}
	P_{RII} = 	\int dE \int \frac{d\sigma_{E}}{dp} W(p) \, dp,
	\label{pale}
	\end{equation}
	\noindent where $E$ denotes the energy of the recollisionally emitted electron and $d\sigma_{E}/dp$ is a cross section for recollision-induced ionization with energy transfer $E$. Introducing the full differential recollision-induced ionization cross section as $d\sigma_{ion}/dp \equiv \int dE\, (d\sigma_{E}/dp)$ we arrive at similar expression to eq.~(\ref{full_excitation_cross_sec}):
	\begin{equation}
	P_{RII} = \int \left( \cfrac{d\sigma_{ion}}{dp}\right)W(p) \, dp,
	\end{equation}	
	
	The full differential recollision-induced ionization cross section $d\sigma_{ion}/dp$ can be regarded as a particular case of the  $d\sigma_{ex}/dp$ because ionization can be treated as an ``excitation to continuum''. Therefore we will use the same assumption on factorization of $d\sigma_{ion}/dp$ into $\tilde{\sigma}_{ion}$ and $f(p)$ terms and rewrite the RII yields in a form
	\begin{equation}
	P_{RII} = \tilde{\sigma}_{ion} \int f(p) W(p) \, dp,
	\end{equation}
	Again, the integral is model-independent and the corresponding RII yields ratio reads then:
	\begin{equation}
	\frac{{P}_{RII}^{3\text{E}}}{{P}_{RII}^{2\text{E}}} \simeq  \frac{\tilde{\sigma}_{ion}^{3\text{E}}}{\tilde{\sigma}_{ion}^{2\text{E}}}.
	\label{RII_ratio}
	\end{equation}

	\subsection{\label{par:eval}Cross section evaluation.} 
	Let us now explain the difference
	of one order of magnitude between ionization yields for the three- and two- electron
	models and validate the assumption (\ref{factorization_p}). For that purpose we will use a simple analysis in Born approximation inspired by Landau and Lifshitz \cite{Landau3quantum}. As we have shown the model-dependence of TDI yield is incorporated in the full differential excitation cross section $d\sigma_{ex}/dp$. Starting with the basic expression for $d\sigma_{ex}$ from the theory of non-elastic collisions (see paragraph 148 of \cite{Landau3quantum}) we write:
	\begin{equation}
	\frac{d\sigma_{ex}}{dp}=  \int_0^p\frac{8\pi}{p^2} \sum_n |\langle np'|U_{ee}|0p\rangle|^2 \frac{dq}{q^3}  \equiv \int d\sigma_{ex}^{(q)},
	\label{gen}
	\end{equation}
	\noindent with $U_{ee}$ being a term in the potential describing the interaction between electrons, $|0\rangle$ and $|n\rangle$ denote the ground and the $n$-th excited state of the parent ion, while $|p\rangle$ and $|p'\rangle$, with $q=p'-p$, denote rescattering electron momenta before and after interaction with ion. Within this notation, the assumption (\ref{factorization_p}) can be expressed as 
	
	\begin{equation}
	\cfrac{d\sigma_{ex}^{(q)}}{dq}=\tilde{\sigma}_{ex} h(q,p), \quad f(p) = \int_0^p  h(p,q)\, dq.
	\label{factorization_q}
	\end{equation}

	In our models $U_{ee}$ reads:
	\begin{equation}
	U_{ee}=\sum_a \frac{q_{ee}^2}{\sqrt{(r-r_a)^2 + r r_a + \epsilon^2}},
	\label{uee}
	\end{equation} where $q_{ee}$ and $\epsilon$ are parameters of the model. Coordinates of the incident and bound electrons are denoted by $r$ and $r_a$, respectively.
	
	First, we calculate the matrix element $\langle p'|U_{ee}|p\rangle$:	
	\begin{multline}
	\langle p'|U_{ee}|p\rangle  = \int e^{-iqr}U_{ee}\, dr =\\  \int e^{-iqr} \sum_a \left( \frac{q_{ee}^2}{\sqrt{(r-r_a)^2 + r r_a + \epsilon^2}} \right) \, dr = \\ q_{ee}^2 \sum_a K_0 \left( q\sqrt{\frac{3}{4}r_a^2+\epsilon^2}\right) e^{-iqr_a/2},
	\label{donkey}
	\end{multline}
	\noindent where $K_0 (q((3/4)r_a^2+\epsilon^2)^{1/2})$ denotes a modified Bessel function of the second kind. It decreases fast with increasing $|r_a|$, thus it is reasonable to expand $e^{-iqr_a} = 1 - iqr_a + O(r_a^2)$ around zero. The constant part of the expansion results in zero matrix element, therefore the approximate expression for the matrix element(\ref{donkey}) reads:
	\begin{multline}
	\langle p'|U_{ee}|p\rangle \simeq -i q_{ee}^2 \sum_a K_0 \left( q\sqrt{\frac{3}{4}r_a^2+\epsilon^2}\right) qr_a/2.
	\label{snake}
	\end{multline}	
	
	After substituting (\ref{snake}) to (\ref{gen}) and applying the rule $\sum_{n \ne 0} |\langle n|f|0 \rangle|^2 = \langle 0|ff^{\dagger}|0 \rangle - |\langle 0|f|0 \rangle |^2$ to the operator (\ref{snake}), and remembering that $|\langle 0|f|0 \rangle |=0$ for the odd $f(r_a)$ one finally gets:	
	\begin{equation}
	d\sigma_{ex}^{(q)} \sim q_{ee}^4 \langle \tilde{d}^2_q \rangle \, dq.
	\label{approx_cross_sec}  
	\end{equation}
	In the above expression we have ignored $p$-, $q$-dependent or constant factors as they are the same for both the models due to similar kinematics of the rescattering electron. Also, a matrix element $\langle \tilde{d}^2_q \rangle $ was introduced:
	\begin{equation}
	\langle \tilde{d}^2_q \rangle \equiv \left\langle  0 \left|  \left[ \sum_a K_0 \left( q\sqrt{\frac{3}{4}r_a^2+\epsilon^2}\right) r_a \right]^2 \right| 0 \right\rangle.
	\label{lama}
	\end{equation}

	It is instructive to compare the obtained result expressed in Eq.~(\ref{lama}) with the analogous one for the full-dimensional case of Coulomb potential $U_{ee}^{\text{Coulomb}}=\sum_a q_{ee}^2/|\textbf{r}-\textbf{r}_a|$. Integration over $\textbf{r}$ gives:
	\begin{equation}
	\langle \textbf{p}'|U_{ee}^{\text{Coulomb}}|\textbf{p}\rangle = \frac{4\pi}{q^2}e^{-i\textbf{q}\textbf{r}_a}.
	\end{equation}
	Therefore one can write:
	\begin{equation}
	d\sigma_{ex}^{(q)\text{Coulomb}} \sim q_{ee}^4 \langle d^2 \rangle \, dq,
	\label{approx_corss_sec_C}
	\end{equation}
	with a matrix element:
	\begin{equation}
	\langle d^2 \rangle \equiv \left\langle  0 \left|  \left[ \sum_a  x_a \right]^2 \right| 0 \right\rangle, \quad x_a = \textbf{r}_a \cdot \frac{\textbf{F}}{|F|}. \label{puma}
	\end{equation}
	\noindent Here $\langle d^2 \rangle$ is a matrix element of square of dipole moment and it does not depend on $q$. So, in the case of Coulomb potential the assumption (\ref{factorization_q},\ref{factorization_p}) is well justified. However, in the case of soft-core potential, the matrix element $\langle \tilde{d}^2_q \rangle$  (see eq.~(\ref{lama})) still depends on transferred momenta $q$.
	
\begin{figure}
		\includegraphics[width=1.0\linewidth]{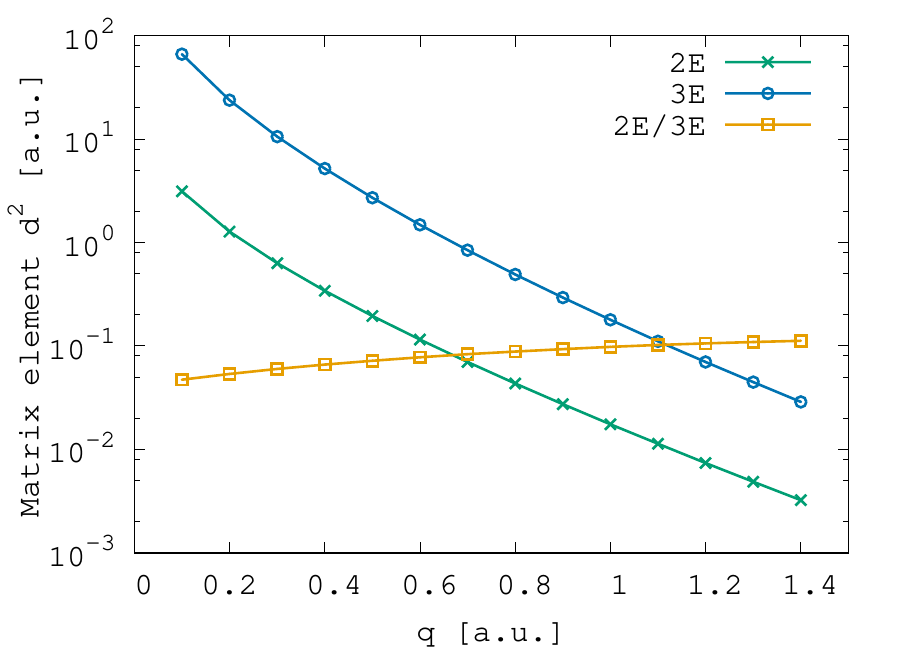}
		\caption{(Color online) Dependencies of matrix element $\langle \tilde{d}^2_q \rangle$ from eq. (\ref{lama}) on rescattering electron momentum difference $q$ for two- and three-electron models along with a ratio of these dependencies. The range of $q$ used on the plot covers possible magnitudes of rescattering electron momentum in the case of external field amplitude $F=0.2$ a.u. 
		}
		\label{fig3}
	\end{figure}

	In Fig.~\ref{fig3} we show $\langle \tilde{d}^2_q \rangle$ as a function of $q$ for both models. In both cases the matrix element decreases monotonically with increasing $q$ in a similar way, the only discrepancy is in the absolute value. Fortunately, the ratio of $\langle \tilde{d}^2_q \rangle$ for different models is nearly $q$-independent. Thus, a reasonable estimation of cross sections ratio can be obtained by putting a constant $q=1$. Then, by defining a modified matrix element of a square of a dipole moment $ \langle \tilde{d}^2 \rangle \equiv \langle \tilde{d}^2_q \rangle|_{q=1}$ such that	
	\begin{multline}
	\langle \tilde{d}^2_q \rangle \simeq \langle \tilde{d}^2 \rangle g(q) = \\ \left\langle  0 \left|  \left[ \sum_a K_0 \left( \sqrt{\frac{3}{4}r_a^2+\epsilon^2}\right) r_a \right]^2 \right| 0 \right\rangle g(q),
	\label{parrot}
	\end{multline}
	\noindent where $g(q)$ is some model-independent function of $q$, one gets analogously to (\ref{approx_corss_sec_C}):
	\begin{equation}
	d\sigma_{ex}^{(q)} \sim q_{ee}^4 \langle \tilde{d}^2 \rangle g(q) \, dq.
	\end{equation}		
	\noindent The differential cross sections $d\sigma_{ex}^{(q)}/dq$ are now clearly separable to the $q$-dependent and the model-dependent $\tilde{\sigma}_{ex} \equiv q_{ee}^4 \langle \tilde{d}^2 \rangle$ factors, proving the assumption (\ref{factorization_q},\ref{factorization_p}). Finally, the ratio of $\tilde{\sigma}_{ex}$ (eqs.~(\ref{TDI_ratio})~and~(\ref{RII_ratio})) reads:	 
	\begin{equation}
	\frac{\tilde{\sigma}_{ex}^{3\text{E}}}{\tilde{\sigma}_{ex}^{2\text{E}}} \simeq \frac{q_{ee}^{4(3E)}}{q_{ee}^{4(2E)}}  \frac{\langle \tilde{d}^2 \rangle^{(3E)}}{\langle \tilde{d}^2 \rangle^{(2E)}} =  \frac{25}{9}\frac{1.78\times 10^{-1}}{1.74\times 10^{-2}} = 28 ,
	\label{triumph}
	\end{equation}	
	\noindent the number that matches well the visible one order of ratio between the ionization yields in the Fig. \ref{fig2}.

	The observed proportionality between $\langle \tilde{d}^2 \rangle^{(3E)}$ and $\langle \tilde{d}^2 \rangle^{(2E)}$ can be explained as follows. Firstly,  observe that $K_0 (((3/4)r_a^2+\epsilon^2)^{1/2})$ decreases fast with increasing $\epsilon$. 
	Secondly, the spatial distribution of the ground state of a single ion is different for both models thus affecting the integration of (\ref{parrot}). In the case of two-electron model, the ground state wavefunction of a single ion is symmetric with respect to exchange of coordinates because it describes a single active electron. As a consequence of this symmetry it has a maximum near the coordinate system's origin. For the three-electron model, however, the ground state wavefunction of a single ion describes two active electrons having the same spin orientations and, thus, it is antisymmetric and has a nodal line along $r_1=r_2$ (that includes nucleus position $r_1=r_2=0$). Therefore, the  wavefunction is broader than its counterpart in the two-electron model. As a benchmark, the values of the average $\langle 0| (\sum_a r_a)^2 |0 \rangle$ evaluated for ground states of single ions in two- and three-electron models are 1.04 a.u. and 3.06 a.u, correspondingly.

	\subsection{The correspondence rule}
	
	The expression (\ref{triumph}) naturally implies the correspondence criterion between different models:
	
	\textit{The TDI efficiency in the knee regime (that is essentially RESI efficiency) is proportional to the ionic ground state diagonal matrix element of a square of a modified dipole moment}.
	
	The form of such a modified dipole moment is defined by the potential term responsible for the interaction between electrons in the system. For example, it is identical to the standard dipole moment operator, providing the Coulomb potential describes the interaction. If the soft-core potential is used instead, the matrix element has a strong dependence on the soft-core parameter~$\epsilon$. In addition, the symmetry of the wavefunction of the ground state of an ion plays a decisive role in calculating the relevant matrix element, as discussed in detail at the end of previous subsection. The symmetry of the wavefunction, in turn, depends on the number of active electrons included in the model. Eventually, such a situation leads to a more efficient recollisional excitation in the models with larger amount of electrons than in the models with fewer electrons.

	There is one more feature that is valid for the cases in which the effective electron-electron charge is introduced. It follows directly from Eqs. (\ref{TDI_ratio}) and (\ref{triumph}): 
	
	\textit{The TDI efficiency is proportional to the forth power of the effective electron-electron charge}.
	
	In the reported case, the models were constructed in such a way that this effective  charge was larger for the three-electron model. 
	
	A straightforward generalization of a differential cross-section notion (\ref{gen}) to the domain of the final free states together with Eq. (\ref{RII_ratio}) implies the same correspondence rule for RII channel. As well as in the case of TDI, the rule is supported by numerical data in Fig. \ref{fig2}.

	\section{\label{sec:conclusions}Conclusions}

	We have performed simulations of ionization dynamics in femtosecond laser pulses for a model atom with the $p^3$ valence shell. Since all the valence electrons have the same spin orientation their wavefunction is antisymmetric with respect to exchange of any pair of electrons. This property allowed us to construct consistent three- and two-active electrons models ready for a direct comparison. Within these models, we investigated how the number of active electrons affects the double ionization yield. In particular, for the laser field amplitudes corresponding to the ``knee'' regime, we have found that the ratio between double ionization yields obtained with three-electron and two-electron models appears to be nearly constant. We have shown that the ratio between double ionization yields may be approximated by the ratio of differential cross sections for recollisional excitation. From the latter the model-dependent elements are easily extracted implying a correspondence rule for double ionization signal magnitudes obtained with different models. 
	
	The increase of excitation cross section while moving from two- to three-electron model agrees well with other trends known from the theory of electronic scattering on atoms (ions) \cite{Landau3quantum}: (i) proportionality of elastic scattering cross section to square of the number of electrons in the target and (ii) linear dependence of Rutherford-type inelastic scattering ($qr_a \gg 1$) on the number of electrons.
	
	\section{Acknowledgements}
	A support by  PL-Grid Infrastructure is acknowledged.
	This work was realized under   National Science Centre (Poland) project Symfonia  No. 2016/20/W/ST4/00314. DKE thanks Artur Maksymov for helpful discussions.
	
\appendix

\newcommand{\snum}{A}

\renewcommand{\theequation}{\snum.\arabic{equation}}
\renewcommand{\thefigure}{\snum.\arabic{figure}}
\setcounter{equation}{0}
\setcounter{figure}{0}

\section{\label{App}Model geometry and flux calculation}

\begin{figure}
	\includegraphics[width=1.0\linewidth]{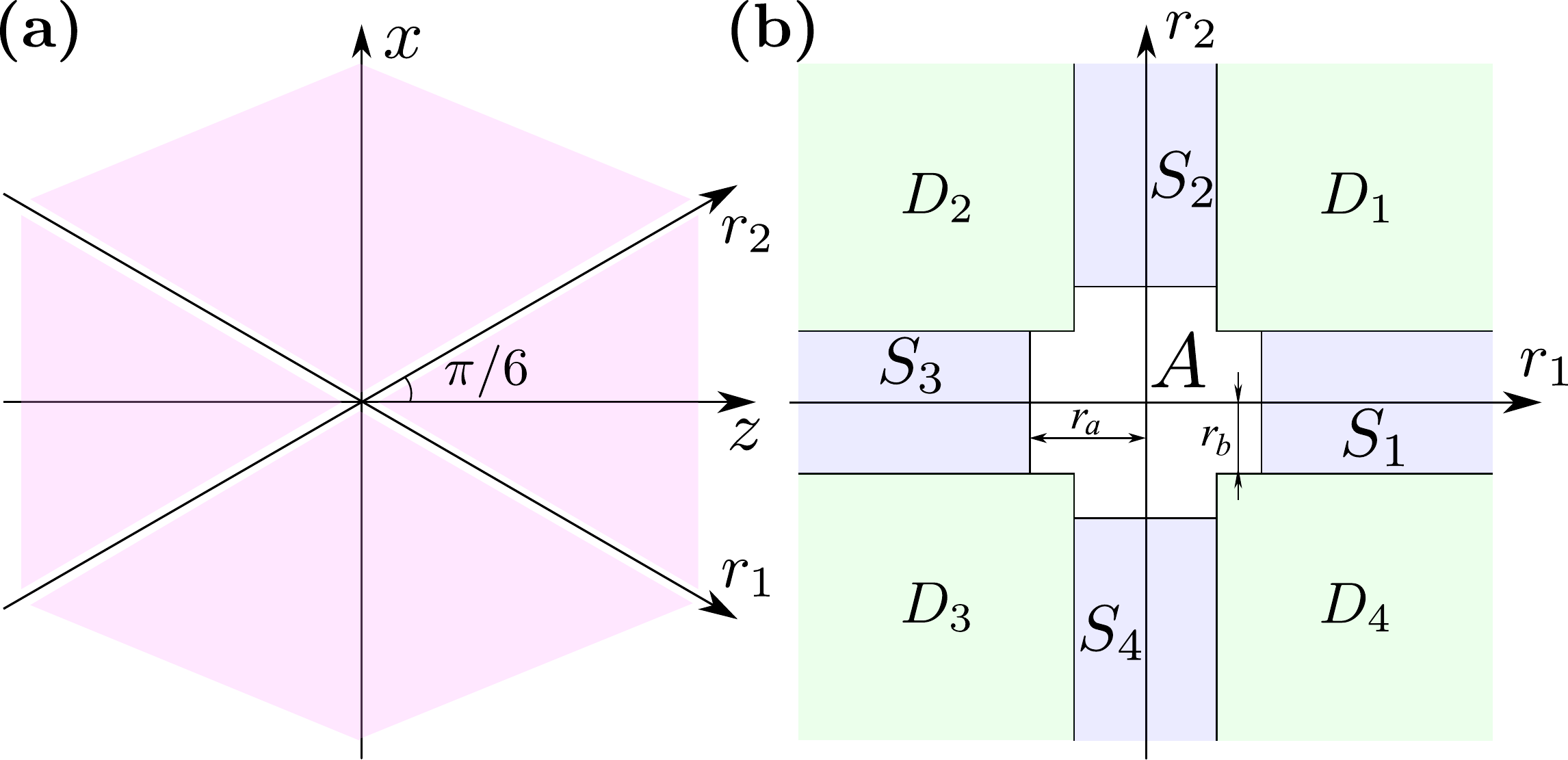}
	\caption{(Color online) (a) Spatial geometry of two-electron model. Electrons propagate along $r_1$ and $r_2$ axes. The field is directed along the $z$-axis. (b) Space division for the two-electron model. The regions correspond to neutral states ($A$), singly ionized states ($S$) and doubly ionized states ($D$). Borders between these regions are defined as $r_a=12.5$ a.u. and $r_b=7$ a.u.
	}
	\label{fig_apx_1}
\end{figure}

\begin{figure}
	\includegraphics[width=1.0\linewidth]{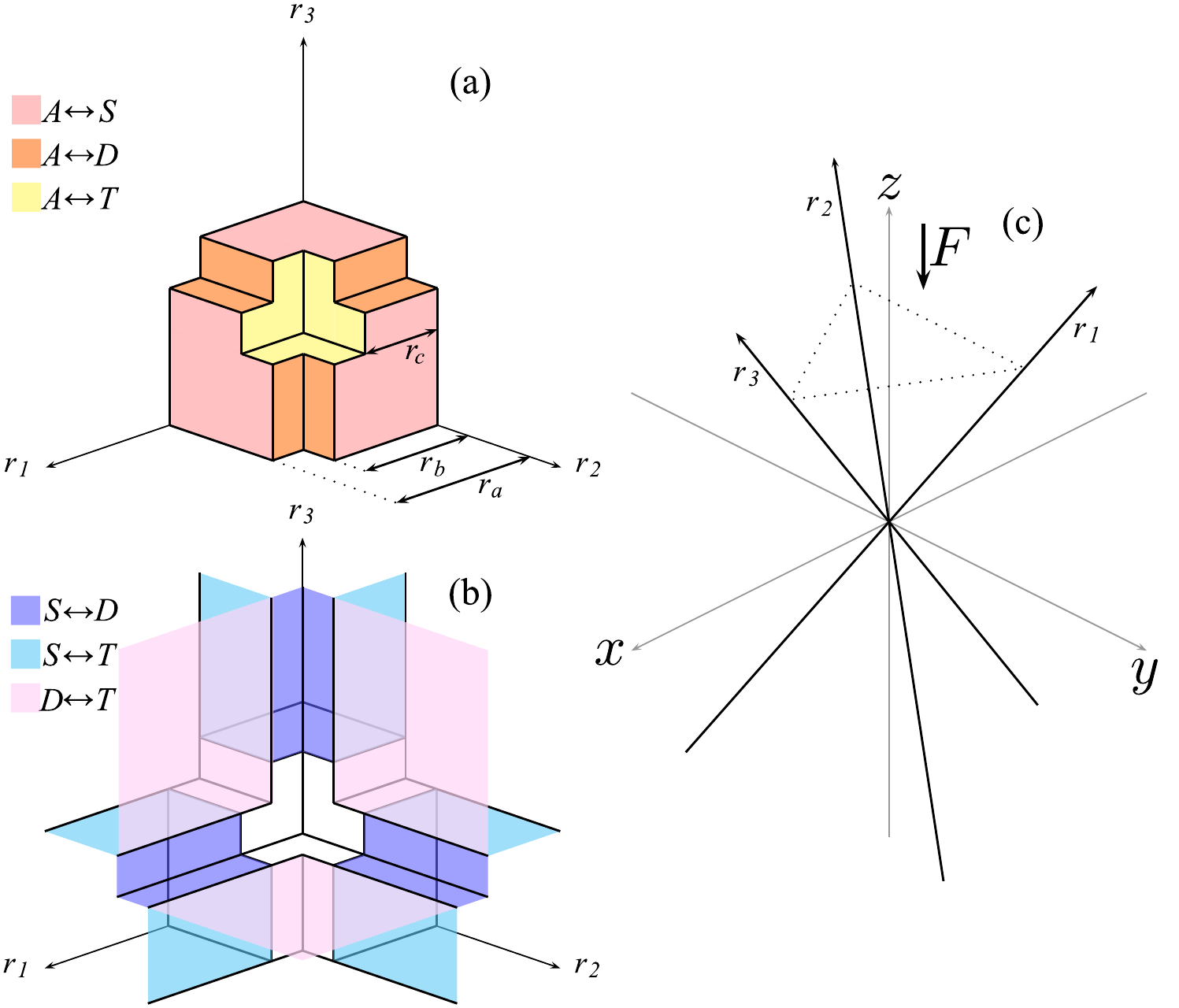}
	\caption{(Color online) (a),(b) Space division for the three-electron model. In color the borders between regions correspond to neutral states ($A$), singly ionized states ($S$), doubly ionized states ($D$) and triply ionized states ($T$). The border distances are $r_a=12$ a.u.,  $r_b=7$ a.u. and $r_c=5$ a.u.(c).Three-electron model. Electrons propagate along $r_1$, $r_2$ and $r_3$ axes. The field polarization direction, ${\vec F}$,   is indicated by the arrow. 
	}
	\label{fig_apx_2}
	\end{figure}

The motion of electrons is restricted along predefined axes. In both two- and three-electron these axes constitute an angle $\gamma=\pi/3$: see mutual position of $r_1$ and $r_2$ in Fig. \ref{fig_apx_1}(a),  ($r_1$ and $r_2$), ($r_3$ and $r_2$), ($r_1$ and $r_3$) in Fig. \ref{fig_apx_2}(c). Laser field direction is chosen to be symmetric in respect to all the electronic axes.

In our algorithm, the total electronic space is divided into regions corresponding to neutral states ($A$), singly ionized states ($S$), doubly ionized states ($D$) and, for three-electron model, triply ionized states ($T$). The populations of these states are calculated 
by integrating the probability fluxes between the corresponding regions \cite{Thiede2018b}. To this end, we assign different spatial regions 
to the different ionization stages and compute the fluxes across the borders. The assignment of the regions has some ambiguities, since it is necessary, for instance, to distinguish a highly excited atomic state with a large excursion of an electron from a singly ionized state where that electron is no longer bound. Nevertheless,  this space separation method is commonly used in both classical and quantum-mechanical  time dependent studies \cite{dundas99,Ruiz05,Becker12}, and provides results
that can be used to deduce trends with external parameters, if the internal assignments of the regions are preserved. The regions and the corresponding borders between them are depicted in Fig. \ref{fig_apx_1}(b) for two-electron model and in Fig. \ref{fig_apx_2}(a),(b) for three-electron model.

The Schr\"odinger equation for the wavefunction $\psi(\textbf{r},t)$ leads, as usual, to the continuity equation
\begin{equation}
\frac{\partial}{\partial t} \rho (\textbf{r},t) + \nabla\cdot \textbf{j}(\textbf{r},t) = 0,
\end{equation}
where the probability density is given by $\rho (\textbf{r},t) = | \psi(\textbf{r},t) |^2$ and 
the probability current by 
\begin{equation}
\textbf{j}(\textbf{r},t) = \Im (\psi^*(\textbf{r},t)\nabla\psi(\textbf{r},t))
\end{equation}
in length gauge or 
by 
\begin{equation}
\textbf{j}(\textbf{r},t) = \Im (\psi^*(\textbf{r},t)\nabla\psi(\textbf{r},t)) - \sqrt{2/3}| \psi(\textbf{r},t) |^2 A(t)
\end{equation}
in velocity gauge with 
vector potential $A(t)$. Changes of the population in region $R \in \mathbb{R}^3$ can be expressed with the application of 
Gauss' theorem as a flux $f_R(t)$ across its borders:
\begin{multline}
\frac{\partial}{\partial t} P_R (\textbf{r},t) = \frac{\partial}{\partial t} \iiint\limits_R |\psi(\textbf{r},t)|^2 \, d^3 \textbf{r} = \\ - \iiint\limits_R \nabla\cdot \textbf{j}(\textbf{r},t)  \, d^3 \textbf{r} = - \iint\limits_{\partial R} \textbf{j}(\textbf{r},t) \cdot d\mathbf{\sigma} \equiv f_R(t),
\label{flux}
\end{multline}
\noindent where $\partial R$ is the border of region $R$ and $d\mathbf{\sigma}$ is the corresponding surface element. 
We assume that the wavefunction decreases sufficiently rapidly as $r\rightarrow \infty$ so that all the above integrals 
converge for any region $R$. Correspondingly, the instantaneous value of the population in region $R$ is given by
\begin{equation}
P_R (\textbf{r},t) = P_R (\textbf{r},0) - \int_0^t f_R(t')\, dt'.
\label{prob}
\end{equation}

The regions for the different states are composed of rectangular domains that are aligned with the coordinate axes, so 
that the boundaries between different regions consist of surfaces parallel to coordinate surfaces.
Following the original proposition \cite{dundas99} we define  the characteristic length $r_a=12.5$ a.u. related to a single ionization (SI) region and $r_b=7$ a.u. related to double ionization (DI) region. For triple ionization we take $r_c=5$ a.u., as suggested by the location of the
triple ionization saddle \cite{Thiede2018b}.  
While these numbers seem somewhat arbitrary,  it may be verified that a reasonable change of the borders leads to small quantitative 
changes of ionization yields obtained only -- the main conclusions about trends as functions of external parameters remain unchanged
if the domains are not modified along the way.

The domains and their boundaries for three-electron model are shown in Fig.~\ref{fig_apx_2}. The region assigned to the atom (label $A$) is the central block in Fig.~\ref{fig_apx_2}(a). Its surface is composed of several segments that stand for transitions to the differently ionized atom: Passing through the three yellow surfaces one electron emitted, so that one enters the single ionization region SI (label $S$). Passing through the orange regions two electrons escape and one enters the double ionization region DI (label $D$). Finally, leaving the atom along the diagonal gives immediate triple ionization TI (label $T$). The notation $i\leftrightarrow j$ used in Fig.~\ref{fig_apx_2} indicates transitions between the different regions. Continuing onwards, there are further boundaries between the ionized states, accounting for transitions between regions SI ($S$) and DI ($D$), for instance (see Fig.~\ref{fig_apx_2}(b)). 

\providecommand{\noopsort}[1]{}\providecommand{\singleletter}[1]{#1}%
%


\end{document}